\documentclass[9pt,shortpaper,twoside,web]{IEEEtran}
\usepackage{generic}
\usepackage{cite}
\usepackage{amsmath,amssymb,amsfonts}
\usepackage{algorithmic}
\usepackage{graphicx}
\usepackage{textcomp}
\usepackage{soul}
\usepackage{multirow,rotating}
\def\BibTeX{{\rm B\kern-.05em{\sc i\kern-.025em b}\kern-.08em
    T\kern-.1667em\lower.7ex\hbox{E}\kern-.125emX}}
\begin{document}
\title{A Radiological Clip Design Using Ultrasound Identification to Improve Localization}
\author{Jenna Cario, Zhengchang Kou, Rita J. Miller, April Dickenson, Christine U. Lee, Michael L. Oelze, \IEEEmembership{Senior Member, IEEE} 
\thanks{ (July 10, 2023)  Many thanks to James Jakub from Mayo Clinic. Research reported in this publication was supported by grants (R21EB030743, R01CA251939, and R01CA273700) from the National Institutes of Health (NIH) and the Cancer Scholars for Translational and Applied Research (C$\star$STAR) Program sponsored by the Cancer Center at Illinois and the Carle Cancer Center under Award Number CST EP082021. The content is solely the responsibility of the authors and does not necessarily represent the official views of the program sponsors. }
\thanks{ J. Cario is with the Beckman Institute for Advanced Science and Technology and the Department of Electrical and Computer Engineering, University of Illinois at Urbana-Champaign, Urbana IL 61801 USA (e-mail: jcario2@illinois.edu). }
\thanks{ Z. Kou is with the Beckman Institute for Advanced Science and Technology, University of Illinois at Urbana-Champaign, Urbana IL 61801 USA (e-mail: zkou2@illinois.edu). }
\thanks{ R. J. Miller is with the Beckman Institute for Advanced Science and Technology and the Department of Electrical and Computer Engineering, University of Illinois at Urbana-Champaign, Urbana IL 61801 USA (e-mail: rjmille@illinois.edu). }
\thanks{ A. Dickenson is with the Breast Imaging Department at Carle Foundation Hospital, Urbana IL 61801 USA (e-mail: April.Dickenson@carle.com). }
\thanks{ C. U. Lee is with the Radiology Department at Mayo Clinic, Rochester MN 55905 USA (e-mail: lee.christine@mayo.edu). }
\thanks{ M. L. Oelze is with the Beckman Institute for Advanced Science and Technology, the Department of Electrical and Computer Engineering, and the Carle Illinois College of Medicine, University of Illinois at Urbana-Champaign, Urbana IL 61801 USA (e-mail: oelze@illinois.edu). }
\thanks{This paper has supplementary downloadable material available at http://ieeexplore.ieee.org, provided by the authors. Included are two animations of ultrasound B-mode data overlaid with results of identification of both one clip and two clips. The total file size is 15.4 MB. }
}

\maketitle
{This work has been submitted to the IEEE for possible publication. Copyright may be transferred without notice, after which this version may no longer be accessible.}

\begin{abstract}
Objective: We demonstrate the use of ultrasound to receive an acoustic signal transmitted from a radiological clip designed from a custom circuit. This signal encodes an identification number and is localized and identified wirelessly by the ultrasound imaging system. 
Methods: We designed and constructed the test platform with a Teensy 4.0 microcontroller core to detect ultrasonic imaging pulses received by a transducer embedded in a phantom, which acted as the radiological clip. Ultrasound identification (USID) signals were generated and transmitted as a result. The phantom and clip were imaged using an ultrasonic array (Philips L7-4) connected to a Verasonics\textsuperscript{\texttrademark} Vantage 128 system operating in pulse inversion (PI) mode. Cross-correlations were performed to localize and identify the code sequences in the PI images. 
Results: USID signals were detected and visualized on B-mode images of the phantoms with up to sub-millimeter localization accuracy. The average detection rate across 4,800 frames of ultrasound data was 93.0\%. Tested ID values exhibited differences in detection rates. 
Conclusion: The USID clip produced identifiable, distinguishable, and localizable signals when imaged. 
Significance: Radiological clips are used to mark breast cancer being treated by neoadjuvant chemotherapy (NAC) via implant in or near treated lesions. As NAC progresses, available marking clips can lose visibility in ultrasound, the imaging modality of choice for monitoring NAC-treated lesions. By transmitting an active signal, more accurate and reliable ultrasound localization of these clips could be achieved and multiple clips with different ID values could be imaged in the same field of view. 

\end{abstract}


\section{Introduction}
\label{sec:introduction}
Breast cancer is among the most common cancers diagnosed in women in the United States, with an estimated 350,000 new cases in 2023 \cite{nih23}. Patients with breast cancer metastasizing to the axillary lymph nodes generally undergo neoadjuvant chemotherapy (NAC) before surgery. NAC may induce a complete pathological response in 40-75\% of node-positive cases \cite{Nguyen17}. Current surgical management after NAC involves targeted axillary dissection, where the sampled positive node is also resected during sentinel lymph node biopsy \cite{bou16,cau16}. As such, markers or clips implanted in the node during needle sampling later confirm that the positive lymph node was resected. In the setting of complete pathological response, however, the marked positive node is difficult to identify by ultrasound, the imaging modality of choice in the axilla, for preoperative localization \cite{hyd19}. 
    

Essentially, during treatment of node-positive breast cancer, two devices are percutaneously implanted: a marker or clip placed in the node at the time of needle sampling \textit{before} starting NAC, and a localizer placed in the same node \textit{after} NAC but before surgery. The majority of the over three dozen breast biopsy clips that are available for use during NAC are metallic, with recent designs employing nitinol to allow for expansion of the marker up to 10 mm once it is deployed from the needle. Despite newer and larger designs, these markers cannot be identified up to 50\% of the time \cite{hyd19,taj23}. This limits the success of ultrasound-guided preoperative radioactive seed localization of the marked node, as the marked target needs to be well seen before deploying a low-dose Iodine-125 seed into the node. Various preoperative techniques to localize the marked nodes are available, including wire localizations, but they can only be performed on the day of surgery \cite{Ram14}. 
    

Several commercially available proprietary localization systems also exist and may be used as an alternative to preoperative strategies requiring same-day placement. However, these systems require clinics to purchase both the localizers and the imaging probes, which are not designed for compatibility with ultrasound systems. For example, the LOCalizer\textsuperscript{\texttrademark} (Hologic, Inc.) is 11 mm long, 2 mm in diameter, and uses radiofrequency (RF) tags and a proprietary probe to identify these RF-enabled tags up to a depth of 6 cm \cite{Malter19}. Radioactive seed localizations, as mentioned above, additionally require a gamma-probe or Geiger counter for detection and must meet the requirements of the Nuclear Regulatory Commission. Another proprietary localization system is the Saviscout surgical guidance system (Merit Medical, South Jordan UT, previously Cianna Medical, Aliso Viejo, CA) which uses radar technology to achieve localization in a surgical setting. This localizer is 1.2 cm long and has two antennae. Lastly, the Magseed (Endomag, Cambridge, UK) requires constraints on surgical equipment when it is near the probe. The costs of these localizing methodologies involve the need for a particular detector, and in some cases, limit the diagnostic quality of other imaging modalities such as MRI, which is often a part of the standard of care in breast imaging in patients with breast cancer.



Color Doppler twinkling has also been used as an ultrasound-compatible approach to identify some markers in preparation for insertion of a localizer. First described in 1996 by Rahmouni et al \cite{rah96}, twinkling is described as rapidly fluctuating colors of an entity, such as kidney stones, during color Doppler ultrasound. The cause of twinkling is still under active investigation, but has been described in the work by Tan, Bi, and Ong \cite{Tan20} with the ULTRACOR\textsuperscript{\textregistered} TWIRL\textsuperscript{\texttrademark} (Bard, Inc.) marking clip as well as other markers \cite{lee21,lee23}. The twinkle artifact appears a few millimeters below the clip and its effect can be enhanced by adjusting system settings. However, to the best of our knowledge, this strategy provides no real-time audio feedback on location and does not enable distinction between multiple tagged areas. 

\section{Our Approach}
\label{sec:background}
In this work, we propose that the identification and localization tasks can be achieved by a single, ultrasound-compatible design, eliminating the need for localizers used specifically during resection and improving reliable visualization during and after NAC. In the proposed scheme, the radiological clip is a powered device, which transmits an ultrasonic signal from a piezoelectric element when imaged by an ultrasound probe. The signal encodes information to achieve ultrasound identification, i.e., USID, and aids in locating the clip itself by adding an additional visual effect in the beamformed data. The imaging system is programmed with additional software or subroutines to detect this visual effect and identify the encoded identification information within the received signal. This information can be further used to provide audio feedback on the probe's proximity to the clip to be used in surgery. 

This clip design requires four major components: a piezoelectric element that can receive and transmit electrical signals through tissue as acoustic pressure waves, a means of triggering a circuit that generates a USID signal, a means of ensuring that the transmitted USID signal does not re-trigger the circuit, and a means of powering the circuit. The overall circuit must be small in size. The small profile allows clips to mark comparatively smaller lesions and be inserted in a minimally-invasive manner. The power source must fit the desired size profile and enable intermittent imaging over several months of NAC, and allow the clip to be imaged and identified afterward during resection. This power source may be either a battery or energy harvested from the imaging probe. Components with low power draw and effective power management are necessary to achieve this requirement. The clip will also need to be able to be encased in a biocompatible shell, conformal coating, or outer layer without major reductions in signal quality or directivity that would inhibit localization. In this work, we evaluate the components needed to achieve the desired functionality in a scalable manner to meet the subsequent size, power, and biocompatibility requirements. However, in this study we do not build an electronic clip at a scale that the clip can be inserted into humans. 

The USID signal itself is also subject to design constraints, and an appropriate USID signal is one that satisfies three requirements. First, it must have an encoding scheme with sufficiently fast decoding so visual and any audio feedback can remain clear and approximately real-time. Second, the signal must be tunable, e.g., multiple USID signals can be generated from a singular design with minimal alteration. Finally, the signal must be long enough to successfully encode information and be detectable, but short enough to be matched with the acquisition rate of an ultrasound imaging system. 

The design we present uses a 64-bit pseudo-noise (PN) code as the USID signal. This scheme is chosen because it can be fetched from a stored code library in read-only memory (ROM) with minimal processing prior to transmit. A 64-bit code transmitted at 4 MHz in tissue has an approximate length of 16 \textmu s. For comparison, the acquisition window used in our experiments is 112 \textmu s. Importantly, PN codes tend, with increasing sequence length, toward having an impulse response at zero lag when auto-correlated. This allows for clear and distinct localization, as a small codebook can be kept by the ultrasound system and decoded quickly via fast Fourier transform (FFT) to cross-correlate the possible codes and the received data. Because PN codes are pseudorandom sequences, there is a possibility that parts of selected codes may cross-correlate strongly with other codes at shifted temporal locations. As such, use of PN codes also necessitates verifying that the autocorrelations of the selected codes are able to be distinguished in magnitude from the cross-correlation of each code in the library with each other code in the library at any temporal shift.

Both the background tissue signal and the systems used to perform imaging can vary greatly. To further increase the signal-to-noise ratio (SNR) of the clips and mitigate the effect of these variables, pulse inversion (PI) imaging is used. Conceptually, pulse inversion imaging arises in ultrasound from modeling the relationship between pressure and density of an acoustic wave in the imaging medium as a Taylor series and considering only the first- and second-order responses, which represent the fundamental and second harmonic, respectively. When the echoes generated by two pulses of identical shape and opposite polarity are summed together, the linear, first-order response vanishes; conversely, the nonlinear, second-order response will be doubled in magnitude. Other, higher order terms that are excluded will follow a similar pattern of even harmonics doubling and odd harmonics vanishing. Typically, PI imaging is conducted at transmit strengths that allow energy to be transferred into higher harmonics to improve image quality over an acquisition using solely the fundamental frequency. However, at lower transmit strengths, tissue response remains approximately linear. As a result, the tissue signal can be greatly attenuated by PI imaging.


The USID circuit only uses the incoming imaging pulses as a trigger to begin signal transmission, so the pulse polarity has no effect on the circuit's response provided both the positive and negative polarity pulses can trigger the circuit at approximately similar locations along their temporal axes. Consequently, the USID signal will appear nonpolar in PI imaging conditions and will resemble a second-order response. When the two acquisitions are summed, the USID signal strength is effectively doubled while the linear tissue signal is removed. PI imaging can therefore be used to isolate the USID signal from the surrounding tissue and the linear tissue signal.

Additionally, if one acquisition is subtracted from the other instead of summed, the nonlinear terms will vanish and the linear tissue signal will be amplified. If these subtracted data are beamformed and presented onscreen, portions that might otherwise be obscured by the USID signal will be visible, and the summed data necessary to identify and localize the USID clip can be processed in the background.

In this paper, we have focused on the basic design elements necessary to test these individual facets in tandem. In particular, we devised a custom printed circuit board (PCB) platform mimicking the intended hardware and transmit/receive functionality of a USID clip, equipped with USID codes designed to increase clip visibility and provide identification. We also created the code necessary for an ultrasound imaging system to be able to decode this information upon receipt. A preliminary version of this work has been reported \cite{car23}.

A similar test configuration was used by Guo, et al. for the purpose of improving visibility of surgical tools, such as catheters, by generating pulses from a small element on the tool's tip \cite{guo14}. In their work, large pulses ($\geq$20 V) were able to be returned to the imaging transducer from the transmitting element. Though our work used a similar test setup and signal flow in hardware, we used lower voltages (\textless 3.3 V) to transmit the USID signals. This was done to emulate receive conditions similar to those as might be observed with transmits from a small, implantable USID device having tight power restrictions and no external wired connections.   


\section{Methods}
\label{sec:methods}

The experimental setup consisted of two parts: 1) the Verasonics research imaging system (Verasonics, Kirkland, WA, USA), its code, and the attached imaging probe, and 2) the circuitry used to simulate the USID clip. These two halves interfaced in an imaging medium via ultrasound. The setup was tested using a tissue-mimicking phantom made of agar and having 70 to 90-\textmu m glass bead scatterers distributed uniformly throughout with spatially random locations. The two 1-mm, epoxy-coated microcrystals (Sonometrics Corp., London, Ontario, Canada) serving as the clip's piezoelectric elements were embedded into the center of this phantom at two depths. The first was approximately 12 mm from the surface of the phantom, and the second was approximately 23 mm from the surface of the phantom. The choice of these depths was based on a previous study that examined 524 clip placements in the breast and found that the range of depths of the clips from the skin surface as observed using ultrasound was between 0.4 to 2.5 cm with an average depth of 1 cm \cite{fal18}.)  In order to visually compare the USID clip signals with a passive commercial clip, a Tumark\textsuperscript{\texttrademark} Eye clip (Hologic Inc., Marlborough, Massachusetts, USA) was included in the phantom between the two crystals. Its depth in the phantom ranged between approximately 15-25 mm from the phantom surface.

\subsection{USID Clip Circuit}

\begin{figure*}[ht]
    \centering
    \includegraphics[width=\textwidth]{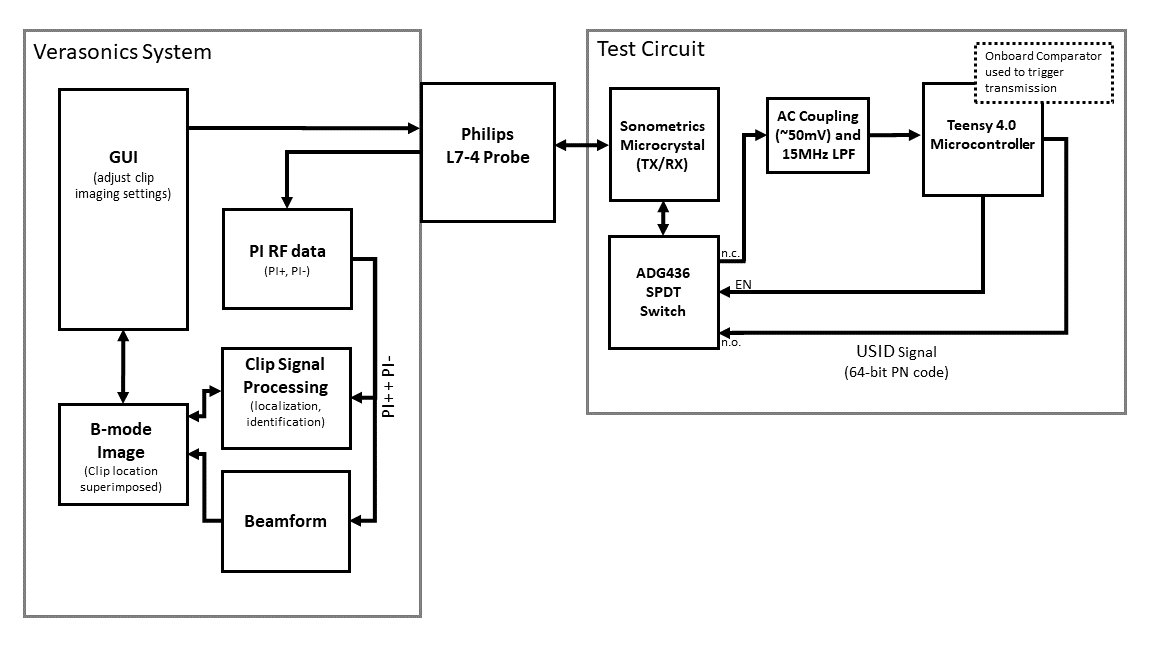}
    \caption{Diagram describing the flow of signals and data through the USID system.}
    \label{fig:flowchartAct}
\end{figure*}

\begin{figure}[ht]
    \centering
    \includegraphics[width=\columnwidth]{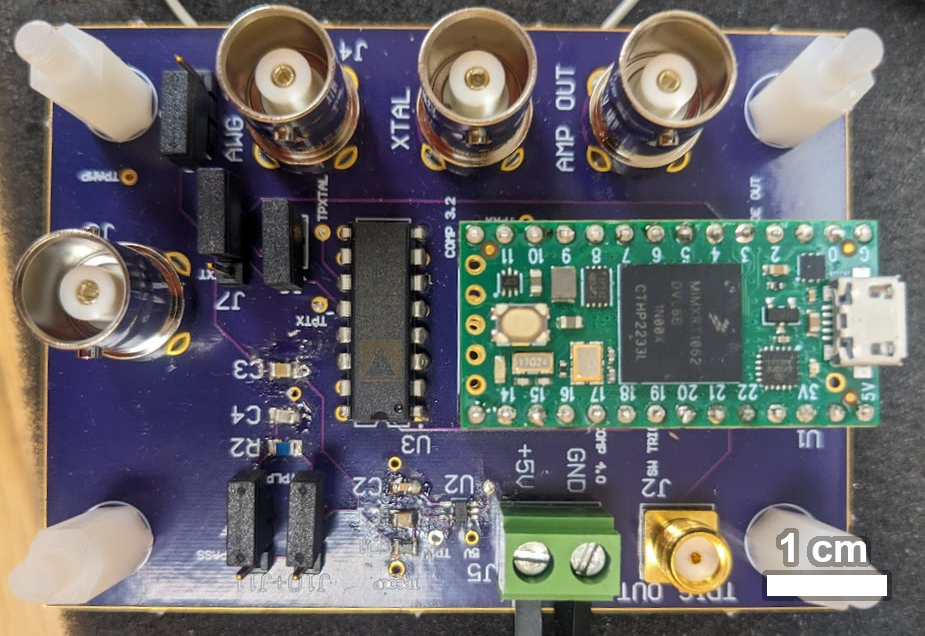}
    \caption{Photograph of PCB used to test USID hardware concept. One PCB was used to control one crystal.}
    \label{fig:PCB}
\end{figure}

Several iterations of USID clip circuitry were tested as the system developed, with the system flow for the experimental configuration described herein illustrated in Fig. \ref{fig:flowchartAct}. At this stage of development, the PCB served as the test bed platform used to evaluate the USID clip concept and generate test USID signals. To this end, this PCB, measuring 79 mm by 58 mm, was used to interface a Teensy 4.0 microcontroller (PJRC, Sherwood, OR, USA) with the microcrystals via BNC cable in order to receive, process, and transmit signals. An ADG436BNZ (Analog Devices, Inc., Wilmington, MA, USA) single-pole, double-throw (SPDT) analog switch was used to route the input and output through a single crystal and to the appropriate Teensy pins. The Teensy 4.0's core processor and its four onboard analog comparators were clocked at 600 MHz. A photograph of this circuit board is shown in Fig. \ref{fig:PCB}. 

The received signal, which could contain an ultrasound imaging pulse, was used as a digital interrupt for the Teensy code. The received signal was filtered through a single-stage RC low-pass filter having a cutoff of 15 MHz to attenuate high-frequency noise. This filtered receive signal generated a software interrupt via one of the comparators, whose expected sensitivity was 52 mV as determined by 64 taps (6 control bits) of the 3.3 V reference voltage. When the interrupt subroutine attached to the comparator was triggered by a voltage change resulting from a received ultrasound pulse, the Teensy transmitted the selected USID code, stored in an array in memory, via a for-loop. This bit-banging solution used \texttt{\textbackslash nop} commands, which execute no operation for a single execution cycle, to create a pulse train with a repetition frequency of 3.90 MHz (approximately matching the center frequency of the imaging probe) and a duty cycle of 33\%. Once generated, the received signal was passed through the SPDT switch to the microcrystal via a wire lead connected to the PCB and into the imaging medium. The transmitted output ranged between 0 and approximately 2.5 V.


\subsection{Verasonics System}

The Verasonics research imaging system leverages MATLAB\textsuperscript{\textregistered} (Natick, MA) to enable user modification of the data-acquisition process. Example code included with the system for PI imaging was used as a template for code to image the USID clip. The code was modified to image with a Philips L7-4 probe (Philips Healthcare, Bothell, WA, USA), a 128-element linear array chosen for the bandwidth overlap shared with the sonomicrocrystals used in the USID circuit. The L7-4 probe fired all elements simultaneously in a single plane wave for each acquisition. Every other acquisition used a pulse that was of opposite polarity to the previous excitation pulse pulse to achieve PI imaging. The pulse repetition frequency (PRF) between two pulses was automatically determined by the imaging system based on the maximum imaging depth. PRF and frame rate were ultimately determined by the total processing time for beamforming and the USID-specific functions added to the code.


A custom delay-and-sum subroutine leveraging GPU capabilities was used for beamforming. A second custom subroutine was created to localize and identify the USID signals (IDs) that might appear in the beamformed data. The chosen localization approach used normalized cross-correlations, which were computed for each of the 128 lines of data, with the potential codes in the codebook. We recorded the actual transmitted codes from the circuit on an oscilloscope, and these recorded waveforms were used as the reference waveforms in the cross correlation. Oscilloscope data of the transmitted codes improves cross-correlation over what would be possible with the "ideal" code by capturing the impulse response of the circuit system. An "ON threshold" for the cross-correlation magnitude served as the lower bound for the presence of a USID signal.

In one operational mode, a specific USID clip can be selected and the imaging system would then look for that code. The USID value of interest was selected on the Verasonics system. The subroutine would determine laterally where this ID's cross-correlation reached its maximum, and if the cross-correlation magnitude was greater than the ON threshold. If so, a detection occurred. On the line containing this maximum response, the axial depth having maximum cross-correlation was used to obtain the location of the USID signal's origin. In this manner, both axial and lateral positions of the USID clip were identified in an image frame. 

Alternately, the operator could select to detect any clip signal and identify and localize it. In this mode, dubbed "freewheeling" mode, the beamformed data were cross-correlated with all possible codes. The code having the highest cross-correlation that crossed the ON threshold was selected as the USID, with otherwise identical steps for lateral and axial localization. In this mode, multiple USID clips could be identified if they were in the field of view.

\subsection{Testing the Clips}

The Verasonics system's user interface has several parameters of interest that were fixed for the tests of the clips and code. These included the time-gain compensation (TGC) left at the default settings and imaging pulse voltage at 13.7 V. This voltage was observed as the lowest step able to trigger both crystals reliably. The ON threshold was set to 0.3 normalized units, where normalization is relative to the geometric mean of the cross-correlated signals' autocorrelations at zero lag. This value was selected based on observations of the system operating in conditions with IDs absent, a single ID present in both selected-ID and freewheeling modes, and two different IDs, with one selected for localization. Test variables included the USID to search for and the depth of the signal (via connection of one of the two embedded crystals). These were used to assess localization accuracy. 

Acquired RF data and beamformed data were saved with relevant system parameters that could be used in post-processing to reconstruct the localization and identification process as needed. In each acquisition, 100 frames of data were gathered. USID data (x- and y- coordinates and ID value) as calculated by the system were also saved. Frames where an ID could not be detected were indicated in this saved data. Localization accuracy over 100 frames was tested for each ID at both possible depths, resulting in 1600 frames of data (8 IDs $\cdot$ 2 depths $\cdot$ 100 frames = 1600 frames). These acquisitions were repeated over $n$ = 3 independent system trials (4800 total frames). Additionally, each frame's processing was independent of every other frame's.

Out-of-plane behavior was also studied by acquiring scans of the 23-mm-deep crystal as the probe was translated across the surface of the phantom with a Daedal micropositioning system (Daedal, Inc., Harrison City, PA, USA). Approximately 100 frames of data for each ID were gathered at 40 \textmu m increments, resulting in a total translation window of approximately 4 mm.

One \textit{in vivo} trial with a female 5-month-old New Zealand White rabbit (Charles River Laboratories, Wilmington, MA) was performed according to a protocol approved by the University of Illinois at Urbana-Champaign Institutional Animal Care and Use Committee (IACUC protocol 21190). The rabbit was anesthetized with 2\% isoflurane, the skin over the mammary fat pad was shaved and disinfected, and a microcrystal was implanted subcutaneously. The crystal was connected to the USID PCB and imaged using the Verasonics system for transmission of IDs 1-4. 100 frames of data were recorded for each ID tested.

\begin{equation}
\label{eq:SNR}
    SNR = 10\log_{10}\left[ \frac{\sigma^{2}_{s+n}}{\sigma^{2}_{n}} -1\right]
\end{equation}

The signal-to-noise ratio (SNR) for each ID at each depth was calculated according to equation \ref{eq:SNR}\cite{liu05}\cite{kan08}, where $\sigma^{2}_{s+n}$ is the variance of the data containing the USID signal plus background noise and $\sigma^{2}_{n}$ is the variance of the background noise without the USID signal. For each ID at both depths, 100 frames of beamformed data were averaged into one image. Variance was calculated laterally along the line of the ground truth signal origin and over the axial range visually determined to contain the USID signal. The background signal was calculated as the variance of an axial line with the same length as that sampled for the signal data but laterally shifted 20 lines toward the center of the image to avoid capturing the fan-shaped tail of the USID signal.  


The capabilities of the "freewheeling mode" were also tested in post-processing, though the code on the Verasonics system for real-time visualization was identical. All 300 frames of data for each ID at both depths were processed using the same ON threshold as was originally used for data collection. The IDs calculated for each detected signal were recorded.

\section{Results}
\label{sec:results}

Under PI imaging conditions, the USID signal was visible as a fan-shaped tail behind the crystal. Signal transmission was verified via oscilloscope to be dependent on the receipt of an imaging pulse through the crystal and on no other factors present in the experimental setup. In some frames, the second of the two pulses would not trigger or would trigger with a longer delay than the first due to jitter in the analog comparator. However, the strength of the transmitted signal tended to be sufficient that these errors did not have an adverse effect on signal visualization or localization. 

To estimate the pressure levels transmitted by the ultrasound array and received by the clip crystal, a 75-\textmu m needle hydrophone (Precision Acoustics, Dorcheseter, UK) was used to measure the pressure levels near the elevational focus of the array, i.e., 25 mm. A peak pressure of 140 kPa was measured in water for an excitation voltage of 13.7 V. Therefore, an approximate maximum of 1.04 mW of power was delivered to each crystal from the imaging probe. The next lower voltage step (11.3 V) corresponded to a peak pressure of approximately 110 kPa in water and was less reliable in triggering the transmission of USID signals from either crystal in the phantom.  

\begin{figure}
    \centering
    \includegraphics[width=\columnwidth]{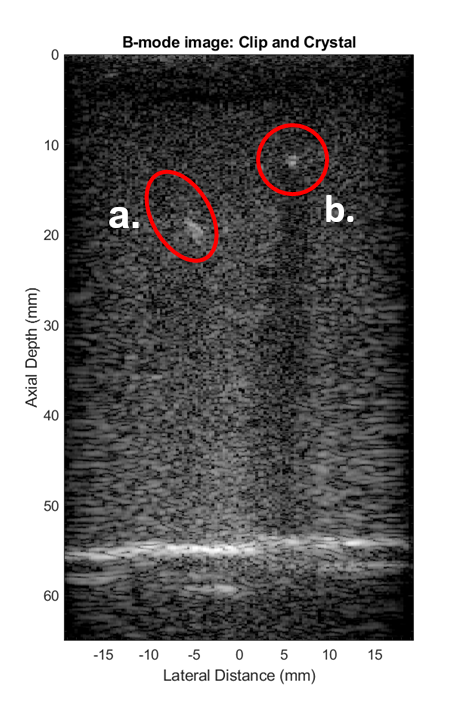}
    \caption{B-mode image, with the Tumark Eye clip (a.) and 12-mm clip (b.) circled.}
    \label{fig:bmodebasic}
\end{figure}

Figure \ref{fig:bmodebasic} shows an annotated view of a B-mode image of the tissue-mimicking phantom, containing the shallower crystal and the metal clip. This view was displayed without using PI imaging and with no USID signals being transmitted. The bottom of the phantom at approximately 55 mm deep appears brighter than either the clip or the crystal.

\begin{table}
        \centering
        \begin{tabular}{p{0.05\linewidth} | p{0.2\linewidth} | p{0.2\linewidth} }
             \textbf{ID} & \textbf{SNR, 12 mm (dB)} & \textbf{SNR, 23 mm (dB)}  \\ \hline
             \rule{0pt}{3ex}1 & 9.115 & 13.797 \\
             2 & 9.160 & 13.260 \\
             3 & 9.980 & 13.912 \\
             4 & 9.480 & 14.020 \\
             5 & 8.331 & 14.324 \\
             6 & 9.079 & 14.132 \\
             7 & 8.088 & 13.786 \\
             8 & 7.963 & 13.393 \\
             \hline
        \end{tabular}
    \caption{SNR value, in dB, for each ID transmitted from both depths.}
    \label{tab:snr}
\end{table}

The SNR (see Table \ref{tab:snr}) calculated for the IDs varied with signal transmission depth, with the 12-mm deep clip providing a signal strength between 3.28 and 6.36 dB less than that achieved for the clip transmitting IDs from 23 mm depth. It should be noted that time-gain compensation (TGC) was enabled for all scans, with increasingly lower sections of depth having higher gain. 


\begin{table}[ht]
    \centering
        \begin{tabular}{p{0.02\linewidth} | p{0.05\linewidth} | p{0.2\linewidth} | p{0.21\linewidth} | p{0.15\linewidth} }
             & \textbf{ID} & \textbf{Mean Error Distance (mm)} & \textbf{Error Distance Variance (mm)} & \textbf{Detection Rate (\%)} \\ \hline
             \multirow{8}{*}{\rotatebox[origin=c]{90}{\textbf{12 mm}}\vspace{12pt}} &  \rule{0pt}{3ex}1  & 1.4452 & 13.7248 & 88\% \\
             & 2 & 0.9410 & 0.0102 & 99\% \\
             & 3 & 0.7724 & 0.0030 & 100\%\\
             & 4 & 0.9313 & 0.0211 & 100\%\\
             & 5 & 1.4461 & 10.0572 & 84\%\\
             & 6 & 0.9620 & 6.8780 & 99\%\\
             & 7 & 3.2267 & 84.0105 & 85\%\\
             & 8\textsuperscript{1} & 1.1607 & 0.0025 & 37\%\\
             \hline
             \multirow{8}{*}{\rotatebox[origin=c]{90}{\textbf{23 mm}}\vspace{12pt}}& \rule{0pt}{3ex}1 & 1.9210 & 16.4677 & 100\%\\
             & 2 & 1.1796 & 5.7010 & 98\%\\
             & 3 & 1.2590 & 0.0946 & 100\%\\
             & 4 & 1.4249 & 5.3247 & 100\%\\
             & 5 & 2.0272 & 3.4786 & 100\%\\
             & 6 & 2.1274 & 15.6610 & 99\%\\
             & 7 & 1.1751 & 1.3214 & 100\%\\
             & 8 & 1.4627 & 2.8166 & 99\%\\ \hline 
             
            \multicolumn{5}{c}{\rule{0pt}{3ex} \textsuperscript{1} ID 8's 12 mm trial 2 had no detections; n = 2 instead of n = 3.}
        \end{tabular}
        
    \caption{Mean and variance of error distance and detection rate for IDs transmitted from both 12- and 23-mm crystals.}
    \label{tab:meanvar}
\end{table}

Key statistics about the localization distance error over the 100 acquired frames in each trial are presented in Table \ref{tab:meanvar}. Averages across the 3 trials were computed as arithmetic means with no weight given based on the number of detections within the trial, so long as at least one successful detection occurred. A mean error in the localization distance of less than 2 mm was observed for most ID-depth pairs across the 3 trials, and some had sub-mm mean error distance. In the trials for the clip at 23 mm depth, no sub-millimeter mean error in localization distances were observed, but all mean errors were still less than 3 mm. Signal detection rate ranged from 98-100\% for the 23-mm trials and from 84-100\% for the 12-mm trials. In the case of ID 8 at 12 mm, one trial failed to detect any present instances of the transmitted signal, resulting in a notably low detection rate. An ON threshold of 0.3 was used for all trials.

\begin{figure}
    \centering
    \includegraphics[width=\columnwidth]{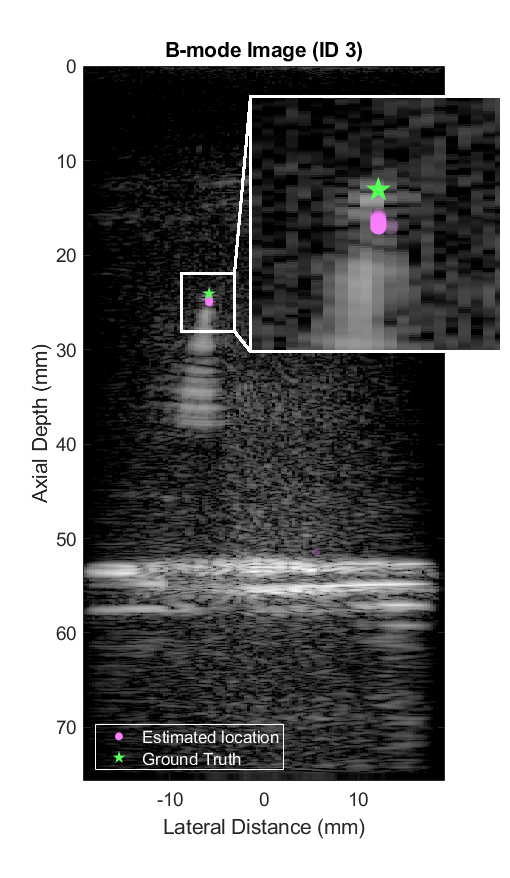}
    \caption{B-mode image overlaid with markers indicating the calculated USID location for ID 3 from the 23-mm crystal. Inset shows region of crystal and localized points in greater detail.}
    \label{fig:IDheatmap}
\end{figure}

Figure \ref{fig:IDheatmap} provides an alternative way to visualize the data from these trials, overlaid on a B-mode image. The ground truth for the 23-mm crystal is marked with a green star. Low-opacity pink circles show the location of the clip as estimated by Verasonics when signal ID 8 was transmitted and queried; areas that Verasonics more commonly attributed as the clip origin for ID 3 over all 100 frames of trial 3 are thus marked with higher opacity to create a heat map. There is a faint pink circle present at 0 mm laterally and axially which, for the purposes of these data, indicated a failed detection. 


\begin{figure}[ht]
    \centering
    \includegraphics[width=\columnwidth]{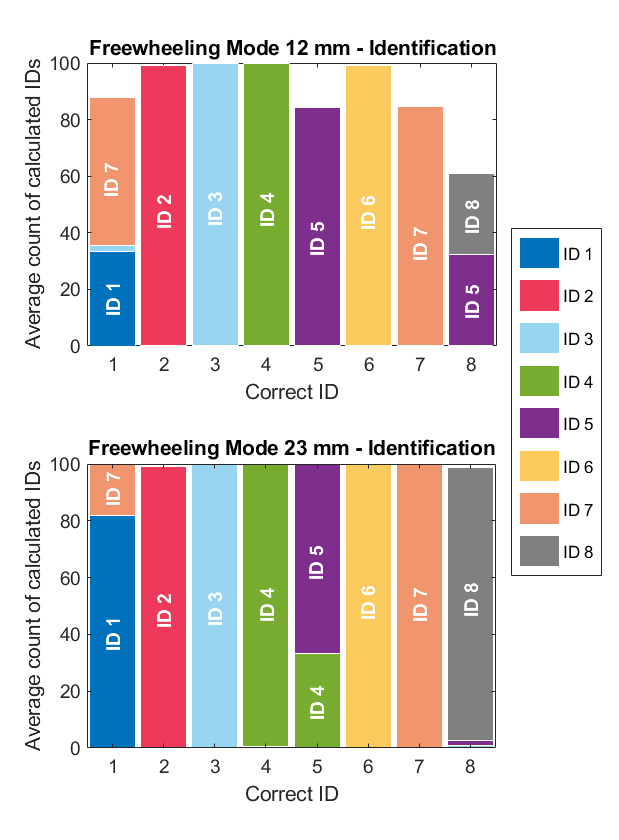}
    \caption{Histograms of calculated ID against the transmitted ID value.}
    \label{fig:freewheel}
\end{figure}

For "freewheeling mode", identification needs to be tested in addition to localization. Figure \ref{fig:freewheel} shows how different IDs were identified. The bins of these histograms are the true value of the transmitted ID, while the histogram counts indicate the calculated ID value, summing up to the number of successful detections per ID, averaged across the three trials. Detection rates for the lower depth and identification accuracy are both higher for the 23-mm-deep transmissions, though ID 5 was confused for ID 4 in about 1/3 of the frames of data. ID 1 also was confused for ID 7 at both depths and more likely to be identified as ID 7 for the 12-mm-deep transmissions. A non-negligible cross-correlation exists between ID 1 and ID 7. ID 8 in the 12-mm-deep transmissions had the lowest rate of detection and was misidentified as ID 5 approximately as often as it was identified correctly. IDs 2, 3, 4, and 6 were identified correctly in nearly all transmissions, regardless of depth.  

As the imaging plane was translated across the 23-mm-deep crystal, the USID signal would begin transmission, persist for approximately 60 frames, and then cease transmission. This corresponds to a translation distance of approximately 2.4 mm (the L7-4 probe has an elevational aperture of 7.5 mm). The epoxy coating encapsulating the 1-mm piezoelectric crystal element was measured as having a diameter of 2.43 mm. When the crystal was mostly in-plane, for approximately 10 frames (0.4 mm), the USID signal appeared brightest and localization was observed to be most accurate. Less optimal alignment of the imaging plane and crystal, which still produced a USID signal, was not as accurate at detection or localization. Across each ID, the locations of turn-on and turn-off of the USID signal were within 5 frames of each other, though the bright USID signal and better localization persisted intermittently beyond 10 frames for some of the IDs.

In the supplementary materials provided with this text, we provide animations showing the performance of the USID localization and identification system in standard mode, querying ID 8 while IDs 8 and 2 transmit from 12- and 27-mm deep, respectively. Subtraction of the two PI acquisitions is used to display the B-mode image animation. A rudimentary version of localization of multiple IDs simultaneously, using the same data, is also provided. Erroneous localization in standard mode is limited within these 100 frames of data, and though multiple IDs may be erroneously identified in a small region due to errors in code, both IDs were captured. No crosstalk was observed between the two crystals.

\begin{figure}
    \centering
    \includegraphics[width=\columnwidth]{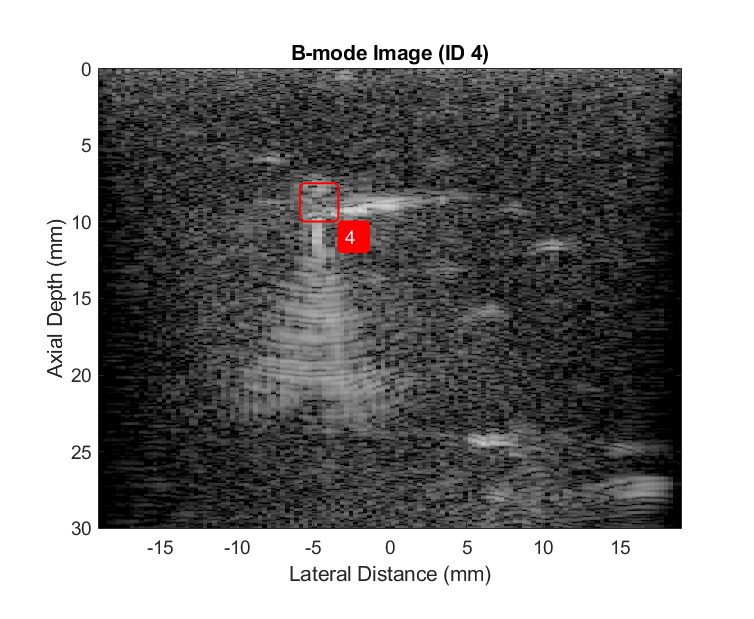}
    \caption{Example frame of \textit{in vivo} data (cropped to a maximum depth of 30 mm) showing USID signal transmission.}
    \label{fig:rabbit} 
\end{figure}

In the \textit{in vivo} data presented in Fig. \ref{fig:rabbit}, we show successful transmission, localization, and post-processing identification of the USID signal. The probe was held at a very shallow angle relative to the skin (less than 45\textdegree) to image the clip. The number of successful triggering events across the four trials was 50 (of 400 frames) and 249 frames had no apparent triggering, which could occur if the image plane was not on the clip. In the remaining frames, triggering occurred at least once per frame, but was offset axially by more than a millimeter.

\section{Discussion}
\label{sec:discussion}

High accuracy was observed in localization of the clip signal for nearly all phantom trials. A mean error in localization distance was less than 2 mm for nearly every ID-depth pair across all trials, which suggests that the clip could be localized successfully. This metric of success is based on clinical implications of having errors in localization. To be successful, localization should be able to overlap the clip along its smallest dimension. Given that the clips are intended to be small and inserted with a needle, the goal is to create a clip design with a smallest dimension no larger than 2 mm. 

Variance in localization error is an indicator of localization stability, as each localization calculation is dependent only on a single frame of data. High variance would suggest a high degree of change in coordinates over the queried frames. This was not observed to be the case for a majority of the data presented. Even in cases where detection rate was low, frames with successful detection clustered about a single point, even if that point was slightly shifted from the ground truth clip position. Because localization calculation was independent from frame to frame, this suggests that signal tracking in a more motion-prone environment with similar background and contrast is feasible, within the limits of pulse inversion's acquisition considerations. The variance in localization of ID 7 at 12 mm depth was the most notable outlier. In trial two of three, 15 of the 100 frames localized to the same point approximately 45 mm distant from the true clip location, while the remaining 85 detections were within 1.15 mm of the true clip location. The difference in these two error-distance groups skewed the variance for that trial to 252 mm. The other trials (i.e., trials one and three) for ID 7 at 12 mm had a variance of 0.0234 and 0.0024, with detection rates of 55\% and 99\%, respectively. The final mean error distance value for ID 7 at 12 mm was likewise also skewed.


The 12-mm-deep crystal was found to be not well aligned in the phantom (i.e., not having maximum response when the probe was placed perpendicular to phantom surface). The probe had to be placed at an angle of approximately 70 degrees relative to the phantom surface elevationally and 84 degrees axially to receive a response with a strength similar to that of the 23-mm-deep crystal imaged at 90 degrees. This misalignment and the subsequent effect on signal quality may have negatively impacted detection rates.


In the \textit{in vivo} validation of the USID concept, subcutaneous implantation of the crystal and its wire lead meant that crystal orientation was always parallel to the skin, increasing the difficulty of imaging and triggering the USID signal transmission despite only thin tissue between the probe and the crystal. Incorrect triggering events occurred frequently, possibly due to tissue motion, signal interference, or strong reverberant echoes. In future \textit{in vivo} experiments, we will image within tumor models that will enable better alignment between the probe and clip while providing additional tissue layers to image through.

The localization and identification procedures reduced the frame rate of the B-mode imaging due to heavy processing loads for cross-correlation. The original frame rate around 30 frames per second recorded without additional algorithms was reduced to between 6 to 12 frames per second. Frame rate started out on the higher end of this range, but reduced to approximately 6 fps over the course of active imaging, suggesting additional code refinement can achieve better frame rates that may be more acceptable in a higher-motion clinical environment. 

The hardware used in this test platform was much larger than would be used in a clinical setting, and requires lead wires for power supply to the board, which is not feasible in practice. Power and biocompatibility considerations for the clip design are also important for clinical translation. Ultrasound power harvesting may be viable, particularly as the necessary USID functionality (including power management) could likely be accomplished within a 500-\textmu W power budget in a 16-nm-process application-specific integrated circuit (ASIC). Such a circuit would also have the advantage of a small footprint, i.e., area less than 1 mm$^{2}$. We expect that the use of a conformal coating such as parylene-C, which is commonly used for other implantable electronics, may also be used for the clip to ensure biocompatibility.  

One shortcoming of this work was the selection of the microcrystal, which had a center frequency of only 1.2 MHz. This off-the-shelf product functioned for this work but was not optimally matched to our needs. We intend to have custom piezoelectric elements with matched impedance, lower directivity, and higher center frequencies for use with future iterations of the clip. We expect the center frequency of these elements will still be somewhat low-frequency, e.g., 4-5 MHz, relative to more typical frequencies for breast imaging, e.g., $\geq$12 MHz. This choice is justified to maintain higher penetration depth and for improved power harvesting. 

Preliminary tests capturing the signal input/output, detection, and memory storage portions of the circuit in 65-nm process IC technology required approximately 250-\textmu m\textsuperscript{2} of chip area. The elements of ASIC design not addressed in the preliminary USID may take the remaining space on the 1-mm-square die. A smaller process such as 16-nm is intended to be used with a final prototype, so the footprint taken by the presently-modeled circuitry would be comparatively smaller and open more chip area for the design elements not yet addressed. Overall device footprint would be influenced by storage capacitor size, crystal size, USID IC size, and connections between these elements and external programming such as the memory address of the ID on a given clip. Capacitance for a storage capacitor can be varied within a standard package size, such as 0603, which has a footprint approximately 1.55 mm by 0.85 mm. The crystal would be no larger than 1 mm by 1 mm. The PCB itself would need to fully encompass each of these devices and would likely be approximately 1.57 mm thick and the thickest component in the design. We predict that these form factors will be sufficient to attain miniaturization for implantation.

\section{Conclusion}
\label{sec:conclusion}

USID signals, which encoded identification information using 64-bit PN codes, were transmitted acoustically from a custom PCB and decoded by the receiving ultrasound system. Cross-correlation was used to successfully localize and identify the signal origins in beamformed B-mode data within 2 mm of their actual position. Localization was accomplished in real time, with frame rates between approximately 6 and 12 fps. The detection rate for the eight tested IDs, tested at depths of 12 mm and 23 mm, was 93.0\%. We envision an implant prototype with a length of no more than 12 mm, and expect that our current level of localization accuracy and detection capability can, through addressing shortcomings we have identified, be improved to a level that is sufficient for both visual localization during NAC as well as for retrieval during surgical procedures. 


\bibliographystyle{IEEEtran}
\bibliography{bib}

\begin{thebibliography}{10}
\providecommand{\url}[1]{#1}
\csname url@samestyle\endcsname
\providecommand{\newblock}{\relax}
\providecommand{\bibinfo}[2]{#2}
\providecommand{\BIBentrySTDinterwordspacing}{\spaceskip=0pt\relax}
\providecommand{\BIBentryALTinterwordstretchfactor}{4}
\providecommand{\BIBentryALTinterwordspacing}{\spaceskip=\fontdimen2\font plus
\BIBentryALTinterwordstretchfactor\fontdimen3\font minus
  \fontdimen4\font\relax}
\providecommand{\BIBforeignlanguage}[2]{{%
\expandafter\ifx\csname l@#1\endcsname\relax
\typeout{** WARNING: IEEEtran.bst: No hyphenation pattern has been}%
\typeout{** loaded for the language `#1'. Using the pattern for}%
\typeout{** the default language instead.}%
\else
\language=\csname l@#1\endcsname
\fi
#2}}
\providecommand{\BIBdecl}{\relax}
\BIBdecl

\bibitem{nih23}
\BIBentryALTinterwordspacing
R.~L. Siegel, K.~D. Miller, N.~S. Wagle, and A.~Jemal, ``Cancer statistics,
  2023,'' \emph{CA: A Cancer Journal for Clinicians}, vol.~73, no.~1, pp.
  17--48, 2023. [Online]. Available:
  \url{https://acsjournals.onlinelibrary.wiley.com/doi/abs/10.3322/caac.21763}
\BIBentrySTDinterwordspacing

\bibitem{Nguyen17}
T.~Nguyen, T.~Hieken, J.~Boughey, and K.~Glazebrook, ``Localizing the clipped
  node in patients with node-positive breast cancer treated with neoadjuvant
  chemotherapy: Early learning experience and challenges.'' \emph{Annals of
  Surgical Oncology}, vol.~24, no.~10, pp. 3011--3016, 2017.

\bibitem{bou16}
J.~C. Boughey, K.~V. Ballman, H.~T. Le-Petross, L.~M. McCall, E.~A. Mittendorf,
  G.~M. Ahrendt, L.~G. Wilke, B.~Taback, E.~C. Feliberti, and K.~K. Hunt,
  ``Identification and {Resection} of {Clipped} {Node} {Decreases} the
  {False}-negative {Rate} of {Sentinel} {Lymph} {Node} {Surgery} in {Patients}
  {Presenting} {With} {Node}-positive {Breast} {Cancer} ({T0}–{T4},
  {N1}–{N2}) {Who} {Receive} {Neoadjuvant} {Chemotherapy}: {Results} {From}
  {ACOSOG} {Z1071} ({Alliance}),'' \emph{Annals of Surgery}, vol. 263, no.~4,
  2016.

\bibitem{cau16}
\BIBentryALTinterwordspacing
A.~S. Caudle, W.~T. Yang, S.~Krishnamurthy, E.~A. Mittendorf, D.~M. Black,
  M.~Z. Gilcrease, I.~Bedrosian, B.~P. Hobbs, S.~M. DeSnyder, R.~F. Hwang,
  B.~E. Adrada, S.~F. Shaitelman, M.~Chavez-MacGregor, B.~D. Smith, R.~P.
  Candelaria, G.~V. Babiera, B.~E. Dogan, L.~Santiago, K.~K. Hunt, and H.~M.
  Kuerer, ``Improved axillary evaluation following neoadjuvant therapy for
  patients with node-positive breast cancer using selective evaluation of
  clipped nodes: Implementation of targeted axillary dissection,''
  \emph{Journal of Clinical Oncology}, vol.~34, no.~10, pp. 1072--1078, 2016,
  pMID: 26811528. [Online]. Available:
  \url{https://doi.org/10.1200/JCO.2015.64.0094}
\BIBentrySTDinterwordspacing

\bibitem{hyd19}
\BIBentryALTinterwordspacing
B.~Hyde, J.~Geske, and C.~Lee, ``Challenges to {I}-125 {Seed} {Localization} of
  {Metastatic} {Axillary} {Lymph} {Nodes} {Following} {Neoadjuvant}
  {Chemotherapy},'' \emph{Journal of Breast Imaging}, vol.~1, no.~3, pp.
  223--229, Aug. 2019, \_eprint:
  https://academic.oup.com/jbi/article-pdf/1/3/223/33501905/wbz032.pdf.
  [Online]. Available: \url{https://doi.org/10.1093/jbi/wbz032}
\BIBentrySTDinterwordspacing

\bibitem{taj23}
\BIBentryALTinterwordspacing
R.~Taj, S.~H. Chung, N.~H. Goldhaber, B.~H. Louie, J.~G. Marganski, N.~S.
  Grewal, Z.~S. Rane, H.~Ojeda-Fournier, A.~Armani, A.~Wallace, and S.~L.
  Blair, ``Localizing positive axillary lymph nodes in breast cancer patients
  post neoadjuvant therapy,'' \emph{Journal of Surgical Research}, vol. 283,
  pp. 288--295, 2023. [Online]. Available:
  \url{https://www.sciencedirect.com/science/article/pii/S0022480422006552}
\BIBentrySTDinterwordspacing

\bibitem{Ram14}
\BIBentryALTinterwordspacing
M.~Ramos, J.~Díez, T.~Ramos, R.~Ruano, M.~Sancho, and J.~González-Orús,
  ``Intraoperative ultrasound in conservative surgery for non-palpable breast
  cancer after neoadjuvant chemotherapy,'' \emph{International Journal of
  Surgery}, vol.~12, no.~6, pp. 572--577, 2014. [Online]. Available:
  \url{https://www.sciencedirect.com/science/article/pii/S1743919114000879}
\BIBentrySTDinterwordspacing

\bibitem{Malter19}
\BIBentryALTinterwordspacing
W.~Malter, J.~Holtschmidt, F.~Thangarajah, P.~Mallmann, B.~Krug, M.~Warm, and
  C.~Eichler, ``First reported use of the faxitron localizer{\texttrademark}
  radiofrequency identification (rfid) system in europe {\textendash} a
  feasibility trial, surgical guide and review for non-palpable breast
  lesions,'' \emph{In Vivo}, vol.~33, no.~5, pp. 1559--1564, 2019. [Online].
  Available: \url{https://iv.iiarjournals.org/content/33/5/1559}
\BIBentrySTDinterwordspacing

\bibitem{rah96}
\BIBentryALTinterwordspacing
A.~Rahmouni, R.~Bargoin, A.~Herment, N.~Bargoin, and N.~Vasile, ``Color doppler
  twinkling artifact in hyperechoic regions.'' \emph{Radiology}, vol. 199,
  no.~1, pp. 269--271, 1996, pMID: 8633158. [Online]. Available:
  \url{https://doi.org/10.1148/radiology.199.1.8633158}
\BIBentrySTDinterwordspacing

\bibitem{Tan20}
M.~P. Tan, Z.~Bi, and E.~M. Ong, ``The ‘twinkle’ artifact — a novel
  method of clip identification to facilitate targeted axillary surgery
  following neoadjuvant chemotherapy in breast cancer patients,''
  \emph{Clinical imaging.}, vol.~68, pp. 36--44, 2020-12.

\bibitem{lee21}
\BIBentryALTinterwordspacing
C.~U. Lee, G.~K. Hesley, S.~Uthamaraj, N.~B. Larson, J.~F. Greenleaf, and M.~W.
  Urban, ``Using ultrasound color doppler twinkling to identify biopsy markers
  in the breast and axilla,'' \emph{Ultrasound in Medicine \& Biology},
  vol.~47, no.~11, pp. 3122--3134, 2021. [Online]. Available:
  \url{https://www.sciencedirect.com/science/article/pii/S0301562921001976}
\BIBentrySTDinterwordspacing

\bibitem{lee23}
\BIBentryALTinterwordspacing
C.~U. Lee, N.~B. Larson, M.~W. Urban, A.~L. Miller, S.~Uthamaraj, M.~A. Piltin,
  J.~W. Jakub, A.~A. Bhatt, J.~F. Greenleaf, and G.~K. Hesley, ``Factors
  associated with ultrasound color doppler twinkling by breast biopsy markers:
  In vitro and ex vivo evaluation of 35 commercially available markers,''
  \emph{American Journal of Roentgenology}, vol. 220, no.~3, pp. 358--370,
  2023, pMID: 36043610. [Online]. Available:
  \url{https://doi.org/10.2214/AJR.22.28107}
\BIBentrySTDinterwordspacing

\bibitem{car23}
\BIBentryALTinterwordspacing
J.~Cario and M.~L. Oelze, ``Design of an electronic marking clip for tracking
  breast cancer lesions through neoadjuvant chemotherapy using ultrasound
  identification,'' vol. 153, no.~3, pp. A357--A357. [Online]. Available:
  \url{https://doi.org/10.1121/10.0019145}
\BIBentrySTDinterwordspacing

\bibitem{guo14}
\BIBentryALTinterwordspacing
X.~Guo, H.-J. Kang, R.~Etienne-Cummings, and E.~M. Boctor, ``Active ultrasound
  pattern injection system (auspis) for interventional tool guidance,''
  \emph{PLOS ONE}, vol.~9, no.~10, pp. 1--13, 10 2014. [Online]. Available:
  \url{https://doi.org/10.1371/journal.pone.0104262}
\BIBentrySTDinterwordspacing

\bibitem{fal18}
\BIBentryALTinterwordspacing
S.~Falcon, R.~J. Weinfurtner, B.~Mooney, and B.~L. Niell, ``Savi scout®
  localization of breast lesions as a practical alternative to wires: Outcomes
  and suggestions for trouble-shooting,'' \emph{Clinical Imaging}, vol.~52, pp.
  280--286, 2018. [Online]. Available:
  \url{https://www.sciencedirect.com/science/article/pii/S0899707118301967}
\BIBentrySTDinterwordspacing

\bibitem{liu05}
J.~Liu and M.~Insana, ``Coded pulse excitation for ultrasonic strain imaging,''
  \emph{IEEE transactions on ultrasonics, ferroelectrics, and frequency control
  a publication of the IEEE Ultrasonics, Ferroelectrics, and Frequency Control
  Society.}, vol.~52, no.~2, 2005.

\bibitem{kan08}
\BIBentryALTinterwordspacing
S.~G. Kanzler and M.~L. Oelze, ``{Improved scatterer size estimation using
  backscatter coefficient measurements with coded excitation and pulse
  compression},'' \emph{The Journal of the Acoustical Society of America}, vol.
  123, no.~6, pp. 4599--4607, 06 2008. [Online]. Available:
  \url{https://doi.org/10.1121/1.2908293}
\BIBentrySTDinterwordspacing

\end{thebibliography}

\end{document}